\documentclass[10pt]{article}

\usepackage{amsmath}
\usepackage{amssymb}
\usepackage{graphicx}
\usepackage{cite}
\usepackage{color}

\usepackage[labelfont=bf,labelsep=period,justification=raggedright]{caption}

\makeatletter
\renewcommand{\@biblabel}[1]{\quad#1.}
\makeatother

\title{\bf Traveling and pinned fronts in bistable reaction-diffusion systems on networks}
\author{Nikos E. Kouvaris$^{1,\ast}$, Hiroshi Kori$^{2}$, Alexander S. Mikhailov$^{1}$}
\date{\small
\begin{flushleft}
$^1$ Department of Physical Chemistry, Fritz Haber Institute of the Max Planck Society, Faradayweg 4-6, D-14195 Berlin, Germany\\
$2$ Department of Information Sciences, Ochanomizu University, Tokyo 112-8610, Japan\\
 $^\ast$ E-mail: nkoub@fhi-berlin.mpg.de
\end{flushleft}}

\begin{document}
\maketitle

\section*{Abstract}

Traveling fronts and stationary localized patterns in bistable reaction-diffusion systems have been broadly studied for classical continuous media and regular lattices. Analogs of such non-equilibrium patterns are also possible in networks. Here, we consider traveling and stationary patterns in bistable one-component systems on random Erd\"os-R\'enyi, scale-free and hierarchical tree networks. As revealed through numerical simulations, traveling fronts exist in network-organized systems. They represent waves of transition from one stable state into another, spreading over the entire network. The fronts can furthermore be pinned, thus forming stationary structures. While pinning of fronts has previously been considered for chains of diffusively coupled bistable elements, the network architecture brings about significant differences. An important role is played by the degree (the number of connections) of a node. For regular trees with a fixed branching factor, the pinning conditions are analytically determined. For large Erd\"os-R\'enyi and scale-free networks, the mean-field theory for stationary patterns is constructed.

\section*{Introduction}

\par
Studies of pattern formation in reaction-diffusion systems far from equilibrium constitute a firmly established research field. Starting from the pioneering work by Turing \cite{turing} and Prigogine \cite{prigogine}, self-organized structures in distributed active media with activator-inhibitor dynamics have been extensively investigated and various non-equilibrium patterns, such as rotating spirals, traveling pulses, propagating fronts or stationary dissipative structures could be observed \cite{kap95,mik94}. Recently, the attention became turned to network analogs of classical reaction-diffusion systems, where the nodes are occupied by active elements and the links represent diffusive connections between them. Such situations are typical for epidemiology where spreading of diseases over transportation networks takes place \cite{col06}. The networks can also be formed by diffusively coupled chemical reactors \cite{kar01} or biological cells \cite{big01}. In distributed ecological systems, they consist of individual habitats with dispersal connections between them \cite{hol08}. Detailed investigations of synchronization phenomena in oscillatory systems \cite{osi07} and of infection spreading over networks \cite{bar08} have been performed. Turing patterns in activator-inhibitor network systems have also been considered \cite{nak10}. 
\par
The analysis of bistable media is of principal importance in the theory of pattern formation in reaction-diffusion systems. Traveling fronts which represent waves of transition from one stable state to another are providing a classical example of self-organized wave patterns; they are also playing an important role in understanding of more complex self-organization behavior in activator-inhibitor systems and excitable media (see, e.g., \cite{mik94,DES09}). The velocity and the profile of a traveling front are uniquely determined by the properties of the medium and do not depend on initial conditions. Depending on the parameters of a medium, either spreading or retreating fronts can generally be found. Stationary fronts, which separate regions with two different stable states, are not characteristic for continuous media; they are found only at special parameter values where a transition from spreading to retreating waves takes place. When discrete systems, formed by chains or fractal structures of diffusively coupled bistable elements, are considered, traveling fronts can however become pinned if diffusion is weak enough, so that stable stationary fronts, which are found within entire parameter regions, may arise \cite{BOO92,ERN93,MIT98a,COS92}. 
\par
In the present study, pattern formation in complex networks formed by diffusively coupled bistable elements is numerically and analytically investigated. Our numerical simulations, performed for random Erd\"os-R\'enyi (ER) or scale-free networks and for irregular trees, reveal a rich variety of time-dependent and stationary patterns. The analogs of spreading and retreating fronts are observed. Furthermore, stationary patterns, localized on subsets of network nodes, are found. To understand such phenomena, an approximate analytical theory for the networks representing regular trees is developed. The theory yields the bifurcation diagram which determines pinning conditions for trees with different branching factors and for different diffusion constants. Its results are used to interpret the behavior found in irregular trees and for ER networks. Statistical properties of stationary patterns in large random networks are moreover analyzed in the framework of the mean-field approximation, which has been originally proposed for spreading-infection  problems \cite{PAS01,col07,COL08} and has also been used in the analysis of Turing patterns on the networks \cite{nak10}.

\subsection*{Bistable systems on networks}

\par
Classical one-component reaction-diffusion systems in continuous media are described by equations of the form

\begin{equation}\label{eq:rdcont}
\dot{u}(\mathbf{x},t) = f(u) + D \nabla^{2} u(\mathbf{x},t)
\end{equation}

\noindent where $u(\mathbf{x},t)$ is the local activator density, function $f(u)$ specifies local bistable dynamics (see Methods) and $D$ is the diffusion coefficient. Depending on the particular context, the activator variable $u$ may represent concentration of a chemical or biological species which amplifies (i.e. auto-catalyzes) its own production. 
\par
In the present study, we consider analogs of the phenomena described by the model \eqref{eq:rdcont}, which are however taking place on networks. In network-organized systems, the activator species occupies the nodes of a network and can be transported over network links to other nodes. The connectivity structure of the network can be described in terms of its adjacency matrix $\mathbf{T}$ whose elements are $T_{ij}=1$, if there is a link connecting the nodes $i$ and $j$ ($i,j=1,...,N$), and $T_{ij}=0$ otherwise. We consider processes in undirected networks, where the adjacency matrix T is symmetric ($T_{ij} = T_{ji}$). Generally, the network analog of system \eqref{eq:rdcont} is given by

\begin{equation}\label{eq:rdnetT}
\dot{u}_i  = f(u_i) + D\sum_{j=1}^{N}\!\left(T_{ij}u_j - T_{ji}u_i \right)
\end{equation}

\noindent where $u_i$ is the amount of activator in network node $i$ and $f(u_{i})$ describes the local bistable dynamics of the activator. The last term in Eq.~\eqref{eq:rdnetT} takes into account diffusive coupling between the nodes. Parameter $D$ characterizes the rate of diffusive transport of the activator over the network links.
\par
Instead of the adjacency matrix, it is convenient to use the Laplacian matrix $\mathbf{L}$ of the network, whose elements are defined as $L_{ij}=T_{ij}-k_i\delta_{ij}$, where $\delta_{ij}=1$ for $i=j$, and $\delta_{ij}=0$ otherwise. In this definition, $k_{i}$ is the degree, or the number of connections, of node $i$ given by $k_i=\sum_jT_{ji}$. In new notations Eq.~\eqref{eq:rdnetT} takes the form
 
\begin{equation}\label{eq:rdnetL}
\dot{u}_i  = f(u_i) + D\sum_{j=1}^{N}\! L_{ij}u_j\,.
\end{equation}

\noindent When the considered network is a lattice, its Laplacian matrix coincides with the finite-difference expression for the Laplacian differential operator after discretization on this lattice. 
\par
A classical example of a one-component system exhibiting bistable dynamics is the Schl\"ogl model \cite{SCHLOEGL72}. This model describes a hypothetical trimolecular chemical reaction which exhibits bistability (see Methods). In the Schl\"ogl model, the nonlinear function $f(u)$ is a cubic polynomial 

\begin{equation}\label{eq:funf}
 f(u)=-\frac{\partial V}{\partial u} =-(u-r_{1})(u-r_{2})(u-r_{3})
\end{equation}

\noindent so that $V(u)$ has one maximum at $r_{2}$ and two minima at $r_{1}$ and $r_{3}$. We have performed numerical simulations and analytical investigations of the reaction-diffusion system \eqref{eq:rdnetL} for different kinds of networks using the Schl\"ogl model.

\section*{Results}

\subsection*{Numerical simulations}

\par
In this section we report the results of numerical simulations of the bistable Schl\"ogl model \eqref{eq:rdnetL} for random ER networks and for trees (the results for random scale-free networks are given in the Supporting Information S1). The ER networks with the mean degree $\langle k \rangle=7$ and sizes $N=150$ or $N=500$ are considered. The trees have several components with different branching factors. The model \eqref{eq:rdnetL} with the parameters $r_1=1$ and $r_3=3$ is chosen; the parameter $r_2$ and the diffusion constant $D$ were varied in the simulations. The parameter $r_2$ was restricted to the interval $1<r_2<2$.

\begin{figure}[t]
\includegraphics{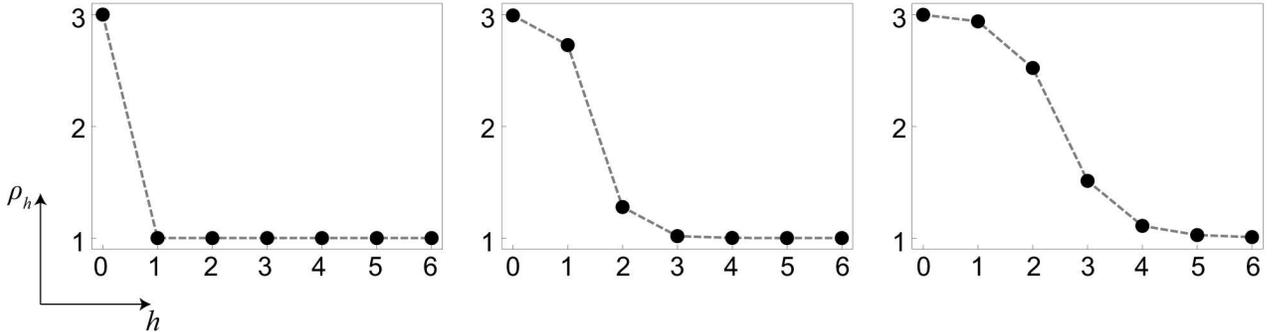}
\caption{{\bf Traveling front in an Erd\"os-R\'enyi network}. The network size is $N=500$ and the mean degree is $\langle k \rangle=7$. Three consequent snapshots of activity patterns at times $t=0,10,21$ are shown. Quantity $\rho_{h}$ is the average value of the activator density $u$ in the subset of network nodes located at distance $h$ from the node which was initially activated. Other parameters are $r_1=1,r_2=1.2,r_2=3$; the diffusion constant is $D=0.1$.}  
\label{fig:ERSpread}
\end{figure}

\par
Traveling activation fronts were observed in ER networks. To initiate such a front, a node at the periphery (with the minimum degree $k$) could be chosen and set into the active state $u=r_3$, whereas all other nodes were in the passive state $u=r_{1}$. This configuration was found to generate a wave of transition from the passive to the active states. The wave spreads from the initially active node to the rest of the system and reaches equidistant nodes, located at the same distance (the shortest path length) from the initial node, at about the same time.
\par 
Front propagation is illustrated in Fig.~\ref{fig:ERSpread}, where the nodes are grouped according to their distance from the first activated node and the average value $\rho_h$ of the activator density $u$ in each group is plotted as a function of the distance $h$. Three snapshots of the traveling front at different times are displayed. As we see, for increasing time the front moves into the subsets of nodes with the larger distances. At the end, all nodes are in the active state $r_{3}$.

\begin{figure}[t]
\includegraphics{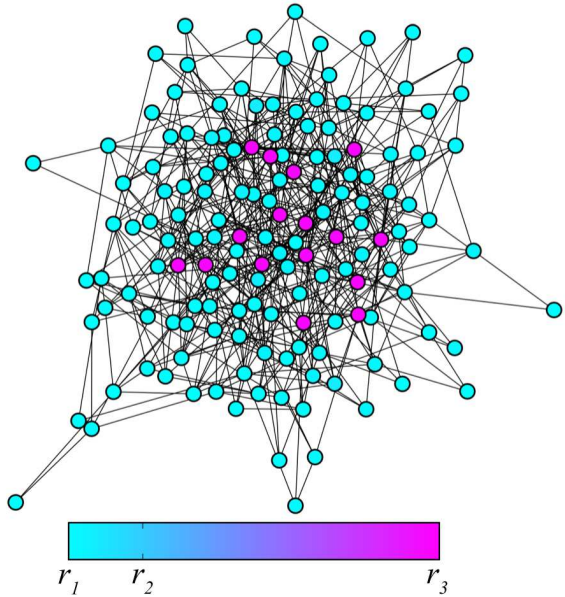}
\caption{{\bf Stationary pattern in an Erd\"os-R\'enyi network}. The network size is $N = 150$ and the mean degree is $\langle k\rangle = 7$. The nodes with higher degrees are located closer to the center. The nodes are colored according to their activation level, as indicated in the bar. The other parameters are $r_1=1$, $r_2=1.4$ and $r_3=3$; the diffusion constant is $D=0.01$.}  
\label{fig:ERstationaryNeato}
\end{figure}

\par
Not all initial conditions lead, however, to spreading fronts. If for example, for the same model parameters as in Fig.~\ref{fig:ERSpread}, a hub node was initially activated, a spreading activation front could not be produced. Retreating fronts were found at these parameter values if the initial activation was set in a few neighbor nodes with large degrees. Under weak diffusive coupling, stationary localized patterns were furthermore observed. If the initial activation was set on the nodes with moderate degrees, the activation could neither spread nor retreat, thus staying as a stationary localized structure. On the other hand, traveling fronts could also become pinned when some nodes were reached, so that the activation could not spread over the entire network and stationary patterns with coexistence of the two states were established. 
\par
\begin{figure}[t]
\includegraphics{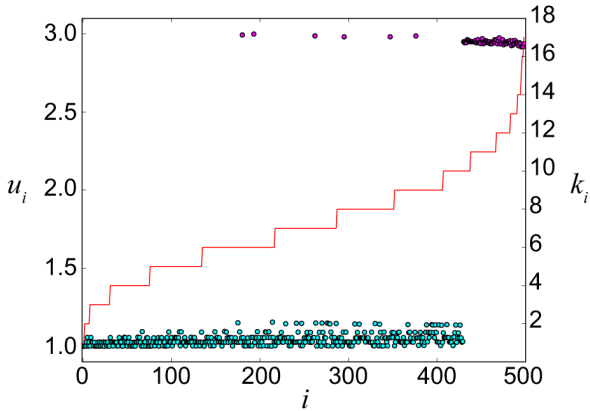}
\caption{{\bf Nodes activation levels for a stationary pattern in an Erd\"os-R\'enyi network}. Dependence of the activation level $u_i$ on the degrees $k_i$ of the nodes $i$ is presented for a stationary pattern in the ER network of size $N = 500$ and mean degree $\langle k \rangle=7$. The nodes are ordered according to their increasing degrees, shown by the stepwise red curve. The other parameters are $r_1=1$, $r_2=1.4$ and $r_3=3$; the diffusion constant is $D=0.01$.} 
\label{fig:ERstationary}
\end{figure}

Degrees of the nodes play an important role in front pinning. In the representation used in Fig.~\ref{fig:ERstationaryNeato}, the nodes with higher degrees lie in the center, whereas the nodes with small degrees are located in the periphery of the network. To produce the stationary pattern shown in this figure, some of the central hub nodes were set into the active state $r_3$, while all other nodes were in the passive state $r_1$. The activation front started to propagate towards the periphery, but the front became pinned and a stationary pattern was formed. Figure~\ref{fig:ERstationary} shows another example of a stationary pattern. Here, we have sorted network nodes according to their degrees, so that the degree of a node becomes higher as its index $i$ is increased (the stepwise red curve indicates the degrees of the nodes). Localization on a subset of the nodes with high degrees is observed.

\par
\begin{figure}[p]
\includegraphics{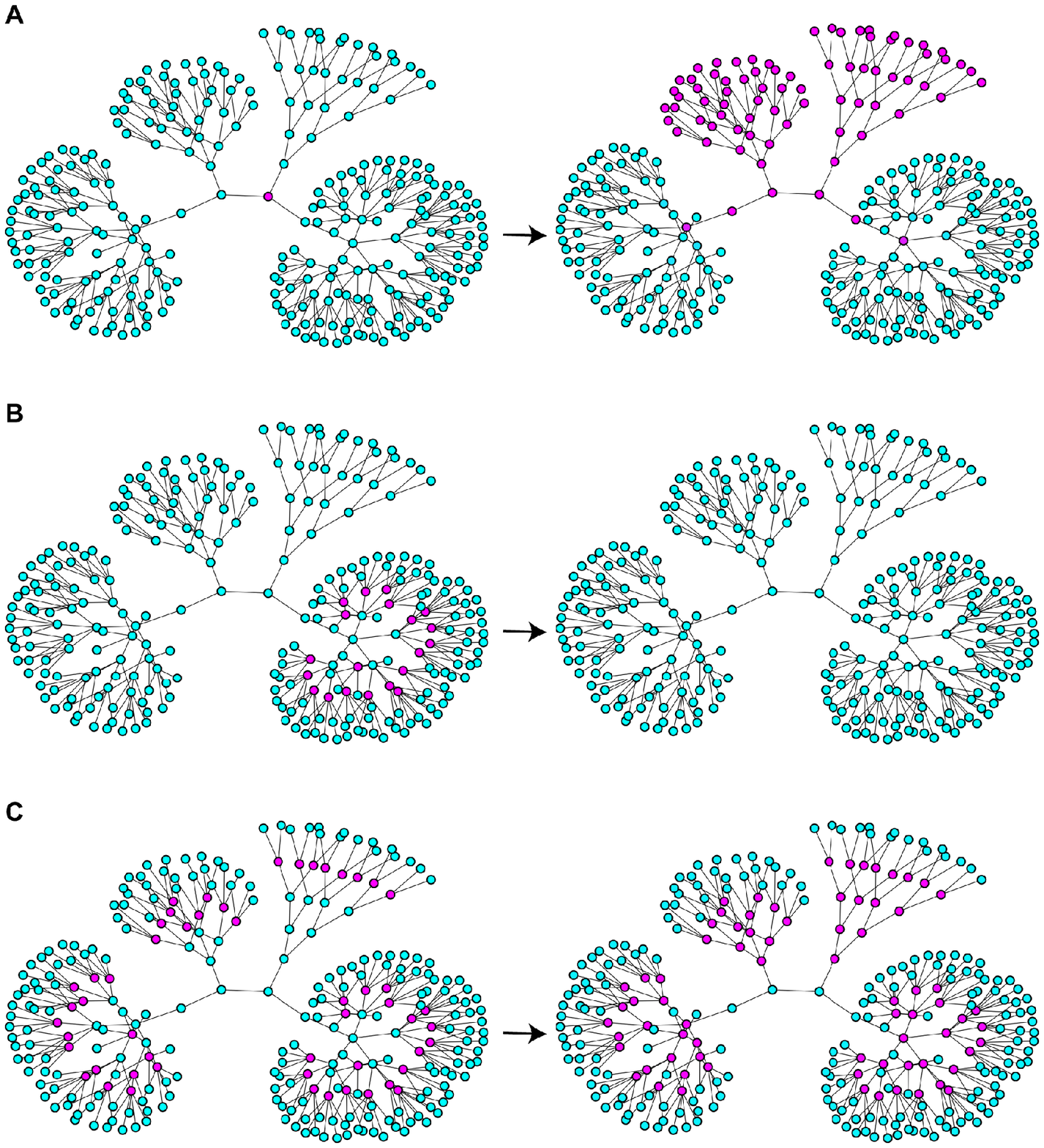}
\caption{{\bf Spreading, retreating and pinning of activation fronts in trees}. A) For $D=0.1$, the fronts spread to the periphery through the nodes with the degrees $k=2,3,4$, while they are pinned at the nodes with the larger degrees. B) For $D=0.35$, the front is retreated from nodes with degree $k=6$. C) For $D=0.03$, the fronts propagate towards the root, but not towards the periphery. In each row, the initial configuration (left) and the final stationary pattern (right) are displayed. The same color coding for node activity as in Fig.~\ref{fig:ERstationaryNeato} is used. Other parameters are $r_1=1$, $r_2=1.4$ and $r_3=3$.}
\label{fig:specialtree}
\end{figure}

The importance of the degrees of the nodes becomes particularly clear when front propagation in the trees with various branching factors is considered (in a tree, the branching factor of a node with degree $k$ is $k-1$). The networks shown in Fig.~\ref{fig:specialtree} consist of the component trees with the branching factors $2,3,4$ and $5$ which are connected at their origins. If the activation is initially applied to the central node, it spreads for $D=0.1$ through the trees with branching factors $2$ and $3$, but cannot propagate through the trees with higher branching factors (Fig.~\ref{fig:specialtree}A). If we choose a larger diffusion constant $D=0.35$ and apply activation to a subset of nodes inside the tree with the branching factor $5$, the activation retreats and dies out (Fig.~\ref{fig:specialtree}B). When diffusion is weak ($D=0.03$), the application of activation inside the component trees leads to its spreading towards the roots of the trees. The activation cannot however propagate further and pinned stationary structures are formed (Fig.~\ref{fig:specialtree}C). 
\par
Thus, we see that both traveling fronts and pinned stationary structures can be observed in the networks. Our numerical simulations suggest that degrees of the nodes (and the related branching factors in the trees) should play an important role in such phenomena. The observed behavior is however complex and seems to depend on the architecture of the networks and on how the initial activation was applied. Below, it is analytically investigated for regular trees with fixed branching factors. The approximate mean-field description for stationary patterns in large random networks is moreover constructed. Using analytical results, complex behavior observed in numerical simulations can be understood. 

\subsection*{Front dynamics in regular trees}

\par
Let us consider the model \eqref{eq:rdnetL} for a regular tree with the branching factor $k-1$. In such a tree, all nodes, lying at the same distance $l$ from the origin, can be grouped into a single shell and front propagation along the sequence of the shells $l = 1,2,3,...$ can be studied. Suppose that we have taken a node which belongs to the shell $l$. This node should be diffusively coupled to $k-1$ nodes in the next shell $l+1$ and to just one node in the previous shell $l-1$. Introducing the activation level $u_l$ in the shell $l$, the evolution of the activator distribution on the tree can therefore be described by the equation

\begin{equation}
\dot{u}_{l} = f(u_{l}) + D (u_{l-1}-u_{l}) + D(k-1)(u_{l+1}-u_{l})\,.
\label{eq:kchain}
\end{equation}

\par
Note that for $k=2$, Eq.~\eqref{eq:kchain} describes front propagation in a one-dimensional chain of coupled bistable elements. Propagation failure and pinning of fronts in chains of bistable elements have been previously investigated \cite{BOO92,ERN93,MIT98a}. The approximate analytical theory for front pinning in the trees, which is presented below, represents an extension of the respective theory for the chains \cite{MIT98a}. Note furthermore that model \eqref{eq:kchain} can be formally considered for any values of $k>2$ of the parameter $k$ (but actual trees correspond only to the integer values of this parameter).
\par 
Comparing the situations for the chains of coupled single elements and for coupled shells in a tree (Eq.~\eqref{eq:kchain}), an important difference should be stressed. In a chain, both propagation directions (left or right) are equivalent, because of the chain symmetry. In contrast to this, an activation front propagating from the root to the periphery of a tree is physically different from the front propagating in the opposite direction, i.e. towards the tree root. As we shall soon see, one of such fronts can be spreading while the other can be pinned or retreating for the same set of model parameters. 
\par
The approximate analytical theory of front pinning can be constructed (cf. \cite{MIT98a}) if diffusion is weak and the fronts are very narrow. A pinned front is found by setting $\dot{u}_{l}=0$ in Eq.~\eqref{eq:kchain}, so that we get

\begin{equation}
f(u_{l}) + D(u_{l-1}-u_{l}) + D(k-1)(u_{l+1}-u_{l}) = 0\,.
\label{eq:pinning}
\end{equation}

\noindent Suppose that the pinned front is located at the shell $l=m$ and it is so narrow that the nodes in the lower shells $l<m$ are all approximately in the active state $r_3$, whereas the nodes in the higher shells are in the passive state $r_1$. Then, the activation level $u_m$ in the interface $l=m$ should approximately satisfy the condition

\begin{equation}\label{eq:pin2}
g(u_{m}) = f(u_{m}) + D \left[(k-1)r_{1} - ku_{m} + r_{3} \right] = 0\,.
\end{equation}

\begin{figure}[t]
\includegraphics{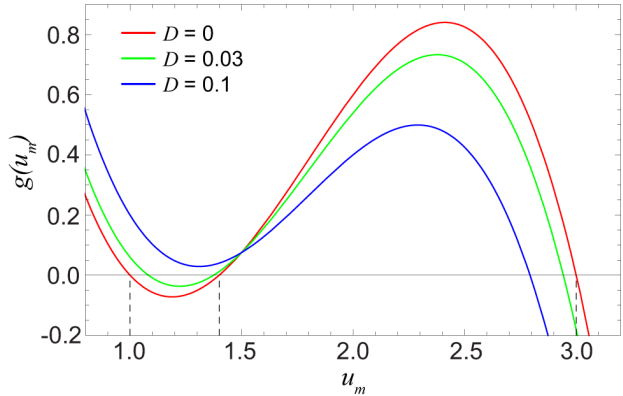}
\caption{{\bf Functions $g(u_{m})$ for three different values of $D$}. The other parameters are $r_1=1$, $r_2=1.4,r_3=3$ and $k=4$.}
\label{fig:nullsD}
\end{figure}

\noindent Thus, the problem becomes reduced to finding the solutions of Eq.~\eqref{eq:pin2}. When $D=0$, we have $g(u_m) = f(u_m$) and, therefore, Eq.~\eqref{eq:pin2} has three roots $u_m = r_1, r_2, r_3$; the front is pinned then. Equation \eqref{eq:pin2} has also three roots if $D$ is small enough (see Fig.~\ref{fig:nullsD} for $D=0.03$). In this situation, the front continues to be pinned. Under further increase of the diffusion constant (see Fig.~\ref{fig:nullsD} for $D=0.1$), the two smaller roots merge and disappear, so that only one (larger) root remains. As previously shown for one-dimensional chains of diffusively coupled elements \cite{MIT98a}, such transition corresponds to the disappearance of pinned stationary fronts.
\par
\begin{figure}[t]
\includegraphics{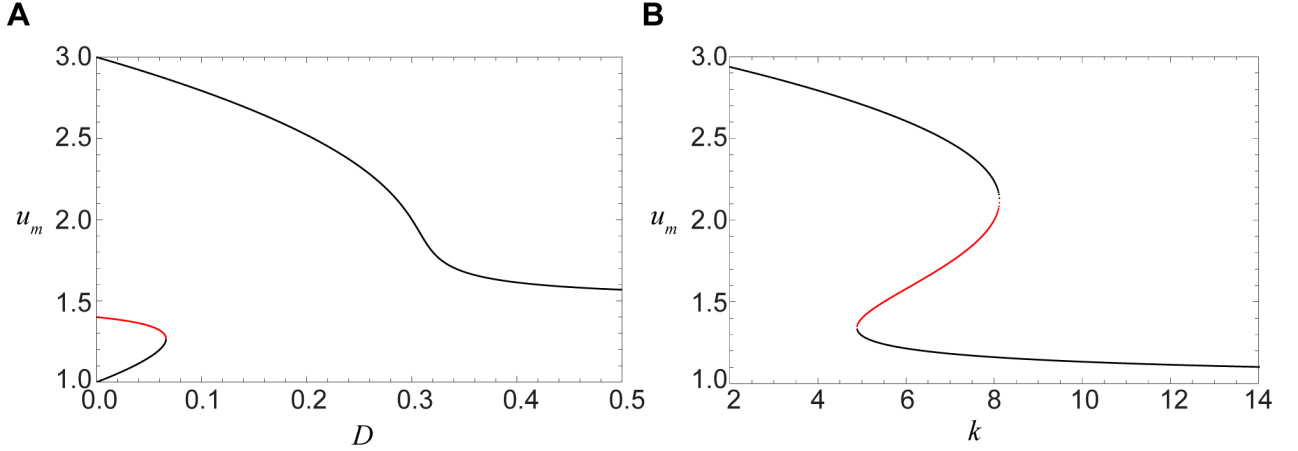}
\caption{{\bf The roots of Eq.~\eqref{eq:pin2}}. The roots $u_{m}$ are plotted as functions (A) of the diffusion constant $D$ for $k=4$ and (B) of the degree $k$ for $D=0.1$. Pinned fronts correspond to red parts of the curves. The model parameters are $r_1=1,r_2=1.4$ and $r_3=3$.}
\label{fig:bifs}
\end{figure}

The transition from pinned to traveling fronts takes place through a saddle-node bifurcation. When $k$ is fixed, the bifurcation occurs when some critical value of $D$ is exceeded (see Fig.~\ref{fig:bifs}A). If the diffusion constant is fixed, pinned fronts are found inside an interval of degrees $k$ (see Fig.~\ref{fig:bifs}B).

\par
Generally, the bifurcation boundary can be determined from the conditions $g(u_{m})=0$ and $g^{\prime}(u_{m})=0$, which can be written in the parametric form as

\begin{eqnarray}\label{eq:sn}
D &=& \frac{\left(2 u_{m} - r_{2} - r_{3}\right)\left(r_{1} - u_{m}\right)^{2}}{r_{1} - r_{3}} \nonumber\\
k &=& \frac{-3 u_{m}^{2} + 2 \left(r_{1}+r_{2}+r_{3}\right)u_{m} - r_{1}r_{2} - r_{1}r_{3} - r_{2}r_{3} }{D}\,.
\end{eqnarray}

\begin{figure}[t!]
\includegraphics{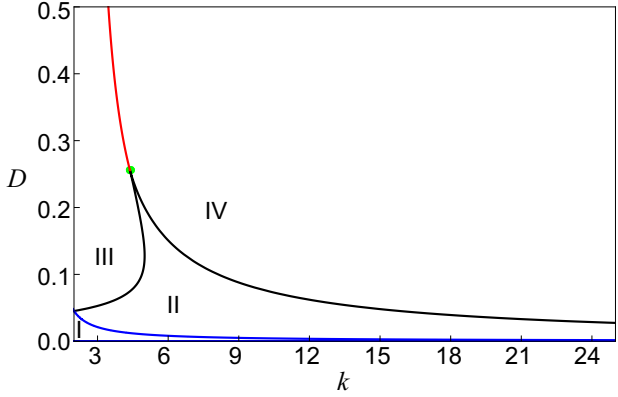}
\caption{{\bf The bifurcation diagram}. Regions with different dynamical regimes are shown in the parametric plane $k-D$. Black curves indicate the saddle-node bifurcations given by the Eq.~\eqref{eq:sn}, while the blue curve stands for the saddle-node bifurcations given by Eq.~\eqref{eq:sn2}. The green dot indicates the cusp point given in Eq.~\eqref{eq:cusp}, the red curve shows the boundary determined by Eq.~\eqref{eq:zerovelocity}. The model parameters are $r_1=1, r_{2}=1.4$ and $r_3=3$.}
\label{fig:bothbifs}
\end{figure}

\noindent Equations \eqref{eq:sn} determine boundaries between regions II and III or II and IV in the bifurcation diagram in Fig.~\ref{fig:bothbifs}. The two boundaries merge in the cusp point, which is defined by the conditions $g(u_{m})=g^{\prime}(u_{m})=g^{\prime\prime}(u_{m})=0$ and is located at

\begin{eqnarray}\label{eq:cusp}
D^{\text{cusp}} &=& \frac{\left(r_{3} + r_{2} - 2r_{1}\right)^{3}}{27\left(r_{3}  - r_{1}\right)}\nonumber\\
k^{\text{cusp}} &=& \frac{r_{1}^{2} + r_{2}^{2} + r_{3}^{2}  - r_{1}r_{2} - r_{1}r_{3} - r_{2}r_{3}}{3D^{\text{cusp}}}\,
\end{eqnarray}

\noindent in the parameter plane. 
\par
Above the cusp point, there should be a boundary line separating regions III and IV. Indeed, fronts propagate in opposite directions in these two regions and, to go from one to another, one needs to cross a line on which the propagation velocity vanishes. This boundary can be identified by using the arguments given below. 

\par
\begin{figure}[t]
\includegraphics{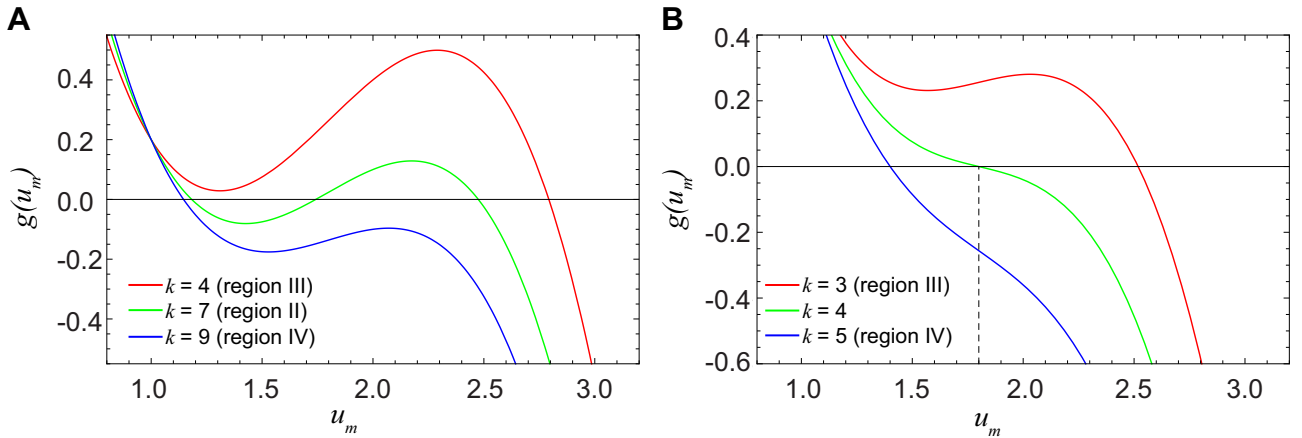}
\caption{{\bf The typical form of functions $g(u_m)$ in different regions of the parameter plane}. Functions $g(u_m)$ are shown below (A, $D = 0.1$) and above (B, $D = 0.32$) the cusp point. The green curve in part (B) corresponds to the boundary between regions III and IV, where the front propagation velocity vanishes. The other parameters are $r_1=1$, $r_2=1.4$ and $r_3=3$.}
\label{fig:nullsK}
\end{figure}

Suppose that the diffusion constant is fixed and $D<D^{\text{cusp}}$. Then the pinned fronts are found inside an interval of degrees $k$, where equation $g(u_m)=0$ has three roots, as in Fig.~\ref{fig:nullsK}A for $k=7$. Outside this interval, equation $g(u_m)=0$ has a single root, which corresponds to spreading fronts if it is close to $r_3$ (Fig.~\ref{fig:nullsK}A for $k=4$) or to retreating fronts if it is close to $r_1$ (Fig.~\ref{fig:nullsK}A for $k=9$). Thus, if we traverse the bifurcation diagram in Fig.~\ref{fig:bothbifs} below $D^\text{cusp}$ by increasing $k$, function $g(u_{m})$ will change its form as shown in Fig.~\ref{fig:nullsK}, having three zeroes within an entire interval of degrees $k$ that corresponds to the pinning region II. When the the diffusion constant is increased and the cusp at $D=D^\text{cusp}$ is approached, such interval shrinks to a point.  If we traverse the bifurcation diagram in Fig.~\ref{fig:bothbifs} above the cusp, the function $g(u_{m})$ changes as shown in Fig.~\ref{fig:nullsK}B. For a given diffusion constant $D$, there is only one degree $k$, such that the function $g(u_{m)}$ has an inflection point coinciding with its zero. The boundary separating regions III and IV is determined by the conditions $g(u_{m})=0$ and $g^{\prime\prime}(u_{m})=0$. In the parameter plane, these conditions yield the curve

\begin{equation}\label{eq:zerovelocity}
D = \frac{\left(r_{1}+r_{2}-2r_{3}\right)\left(r_{2}+r_{3}-2r_{1}\right)\left(r_{1}+r_{3}-2r_{2}\right)}
{9\left[3(r_{3} - r_{1})-\left(r_{3} + r_{2} - 2r_{1}\right)k\right]}\,
\end{equation}

\noindent where the propagation velocity of the fronts is changing its sign. 

\par
The above results refer to the first kind of fronts, where the nodes in the lower shells ($l<m$) of the tree are in the active state $u=r_{3}$ and the nodes in the periphery are in the passive state $u=r_{1}$. A similar analysis can furthermore be performed for the second kind of fronts, where the nodes in the periphery are in the active state and the nodes in the lower shells are in the passive state. Such pinned fronts are again determined by equation \eqref{eq:pin2}, where however the parameters $r_{3}$ and $r_{1}$ should be exchanged. The pinning boundary for them can be obtained from equations \eqref{eq:sn} under the exchange of $r_{1}$ and $r_{3}$. This yields

\begin{eqnarray}\label{eq:sn2}
D &=& \frac{\left(r_1 + r_2 - 2 u_m\right)\left(r_3 - u_m\right)^{2}}{r_1 - r_3} \nonumber\\
k &=& \frac{-3 u_{m}^{2} + 2 \left(r_{1}+r_{2}+r_{3}\right)u_{m} - r_{1}r_{2} - r_{1}r_{3} - r_{2}r_{3} }{D}\,.
\end{eqnarray}

\noindent Fronts of the second kind are pinned for sufficiently weak diffusion, inside region I in the bifurcation diagram in Fig.~\ref{fig:bothbifs}. The boundary of this region is determined in the parametric form by Eqs.~\eqref{eq:sn2}. 
\par
\begin{figure}[t!]
\includegraphics{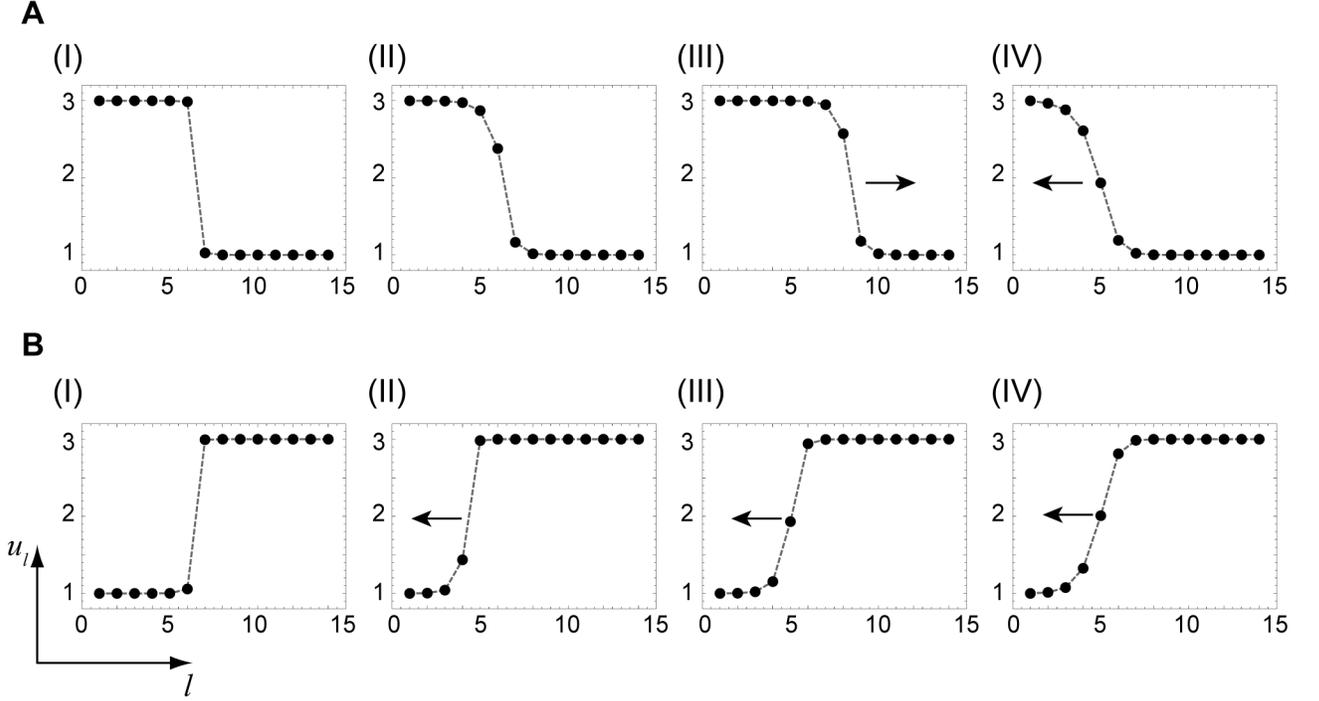}
\caption{{\bf Stationary and traveling fronts in regular trees}. The arrows show the propagation direction for traveling fronts. The labels refer to different regions of the bifurcation diagram in Fig.~\ref{fig:bothbifs}. They correspond to the parameter values (I) $k=3,D=0.01$, (II) $k=6,D=0.03$, (III) $k=3,D=0.1$ and (IV) $k=12,D=0.1$. The other parameters are $r_1=1,r_2=1.4$ and $r_3=3$.}
\label{fig:treefronts}
\end{figure}

Thus, our approximate analysis has allowed us to identify regions in the parameter plane ($k,D$) where the fronts of different kinds are pinned or propagate in specific directions. Predictions of the approximate analytical theory agree well with numerical simulations for regular trees. Figure~\ref{fig:treefronts} shows traveling and pinned fronts found by direct integration of Eq.~\eqref{eq:kchain} in different regions of the parameter plane. For each region, the behavior of two kinds of the fronts, with the activation applied to the nodes of the lower shells ($l\leq6$) or periphery nodes ($l>6$), is illustrated. When the parameters $D$ and $k$ are chosen within region I of the bifurcation diagram, both kinds of fronts are pinned (Figs.~\ref{fig:treefronts}A(I),B(I)). In region II, the front initiated from the tree origin is pinned (Fig.~\ref{fig:treefronts}A(II)), whereas the front initiated in the periphery propagates towards the root (Fig.~\ref{fig:treefronts}B(II)). Activation fronts which propagate in both directions, towards the root and the periphery, are found in region III (Fig.~\ref{fig:treefronts}A(III),B(III)). In region IV, the activation front initiated at the root is retreating (Fig.~\ref{fig:treefronts}A(IV)), whereas the front initiated at the periphery is spreading (Fig.~\ref{fig:treefronts}B(IV)).

\par
In addition to providing examples of the front behavior, Fig.~\ref{fig:treefronts} also allows us to estimate the accuracy of approximations used in the derivation of the bifurcation diagram. In this derivation, we have assumed (similar to Ref. \cite{MIT98a}) that diffusion is weak and the fronts are so narrow that only in a single point the activation level differs from its values $r_1$ and $r_{3}$ in the two uniform stable states. Examining Fig.~\ref{fig:treefronts}, we can notice that this assumption holds well for the lowest diffusion constant $D = 0.01$ in region I, whereas deviations can be already observed for the faster diffusion in regions II, III and IV. Still, the deviations are relatively small and the approximately analytical theory remains applicable. 

\begin{figure}[t!]
\includegraphics{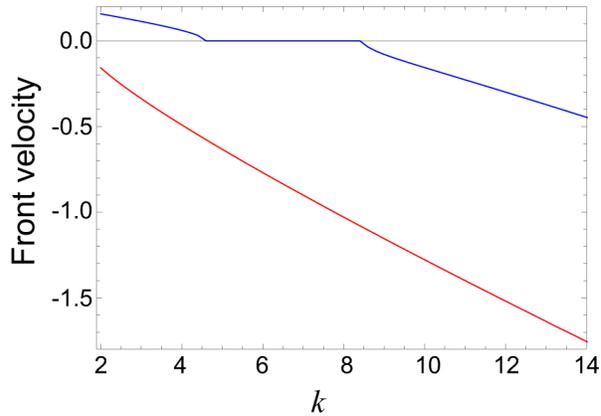}
\caption{{\bf Dependence of the front velocity on node degree}. The blue curve corresponds to the first kind of fronts, shown in  Fig.~\ref{fig:treefronts}A; the red curve is for the second kind of fronts shown in Fig.~\ref{fig:treefronts}B. The diffusion constant is $D=0.1$; the other parameters are $r_1=1,r_2=1.4$ and $r_3=3$.}
\label{fig:frontsvelocity}
\end{figure}

\par
Figure \ref{fig:frontsvelocity} shows the numerically determined propagation velocity of both kinds of activation fronts for different degrees $k$ at the same diffusion constant $D = 0.1$. The blue curve corresponds to the fronts of the first kind, with the activation applied at the tree root. Such front is spreading towards the periphery for small degrees $k$ (region III), is pinned in an interval of the degrees corresponding to region II, and retreats towards the root for the larger values of $k$ (region IV). The red curve displays the propagation velocity for the second kind of fronts, with the activation applied at the periphery. For the chosen value of the diffusion constant, such front is always spreading, i.e. moving towards the root. We can notice that, for the same parameter values, the absolute propagation velocity of the second kind of fronts is always higher than that for the fronts of the first kind. 

\begin{figure}[t]
\includegraphics{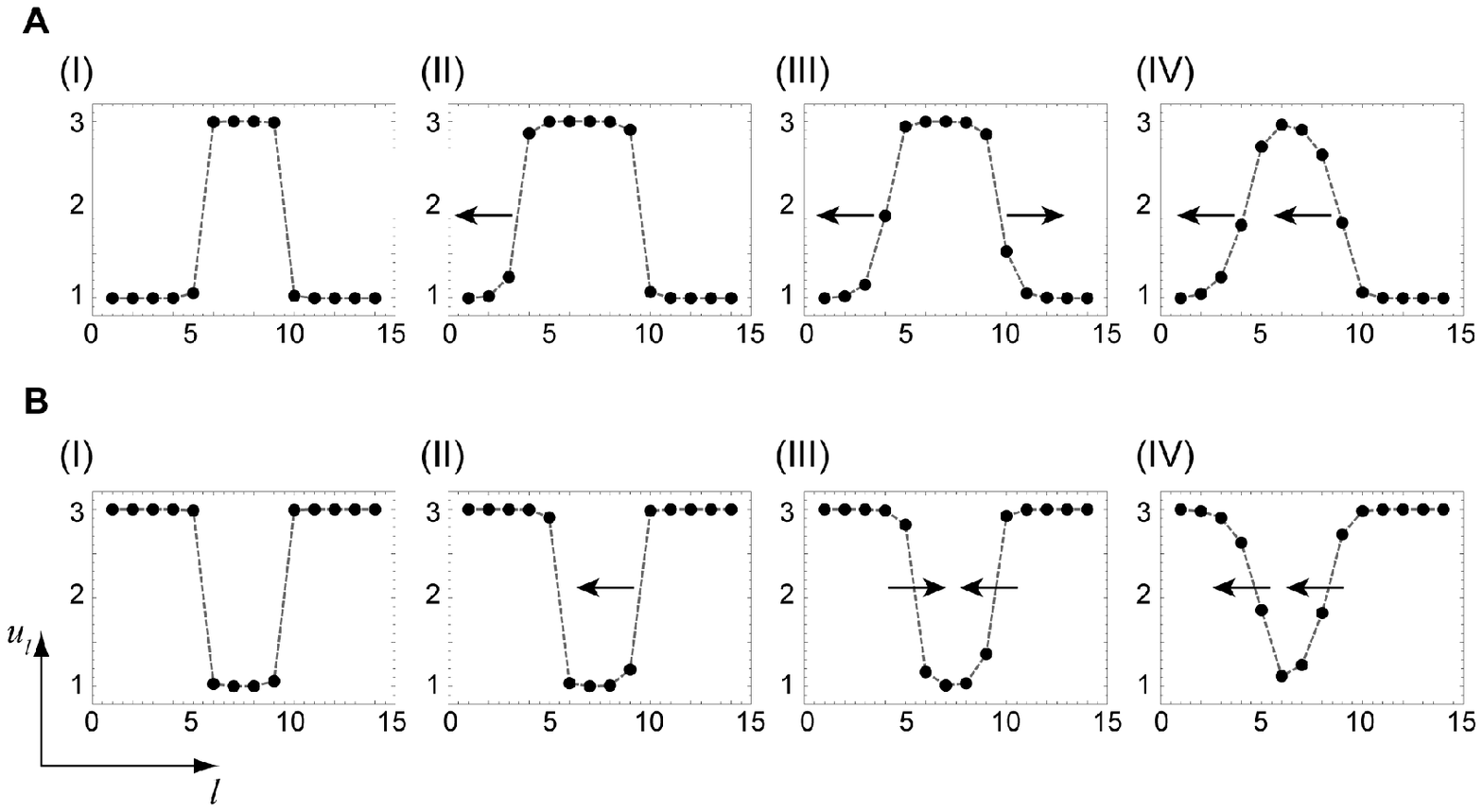}
\caption{{\bf Nonlinear evolution of local perturbations}. The evolution of different local perturbations (A,B) is shown in various regions of the bifurcation diagram. The arrows show the propagation direction. The parameter values are (I) $k=3,D=0.01$, (II) $k=6,D=0.03$, (III) $k=3,D=0.1$ and (IV) $k=12,D=0.1$. The other parameters are $r_1=1,r_2=1.4$ and $r_3=3$.}
\label{fig:treepatterns}
\end{figure}

\par
Using Fig.~\ref{fig:treefronts}, we can consider evolution of various localized perturbations in different parts of the bifurcation diagram (Fig.~\ref{fig:bothbifs}). Inside region I, all fronts are pinned. Therefore, any localized perturbation  (see Fig.~\ref{fig:treepatterns}A(I),B(I)) is frozen in this region. If the activation is locally applied inside region II, it spreads towards the root, but cannot spread in the direction to the periphery (Fig.~\ref{fig:treepatterns}A(II)). On the other hand, if a local ``cold'' region is created in region II on the background of the ``hot'' active state, it shrinks and disappears (Fig.~\ref{fig:treepatterns}B(II)). In region III, local activation spreads in both directions, eventually transferring the entire tree into the hot state (Fig.~\ref{fig:treepatterns}A(III)), whereas the local cold region on the hot background shrinks and disappears (Fig.~\ref{fig:treepatterns} B(III)). An interesting behavior is found in region IV. Here, both kinds of fronts are traveling in the same direction (towards the root), but the velocity of the second of them is higher (cf. Fig.~\ref{fig:frontsvelocity}). Therefore, the hot domain would gradually broaden while traveling in the root direction (Fig.~\ref{fig:treepatterns}A(IV)). The local cold domain (Fig.~\ref{fig:treepatterns}B(IV)) would be however shrinking while traveling in the same direction.
\par
With these results, complex behavior observed in numerical simulations for the trees with varying branching factors (Fig.~\ref{fig:specialtree}) can be understood. In the simulation shown in Fig. \ref{fig:specialtree}A, the diffusion constant was $D=0.1$ and, according to the bifurcation diagram in Fig.~\ref{fig:bothbifs}, the nodes with degrees $k=2,3,4$ should correspond to region III, while the nodes with the higher degrees $k=5,6$ are in the region II. Indeed, we can see in Fig.~\ref{fig:specialtree}A that activation can propagate from the root over the subtrees with the small branching factors, but the front fails to propagate through the subtrees with node degrees $5$ and $6$. In the simulation for $D = 0.35$ shown in Fig.~\ref{fig:specialtree}B, the activation has been initially applied to a group of nodes with degree $k=6$ corresponding to region IV. In accordance with the behavior illustrated in Fig.~\ref{fig:treepatterns}A(IV), such local perturbations broaden while traveling towards the root of the tree, but get pinned and finally disappear. In Fig.~\ref{fig:specialtree}C we have $D=0.03$ and, therefore, we are in region II for all degrees $k$.  According to Fig.~\ref{fig:treepatterns}A(II), local activation in any component tree should spread towards the root, but cannot propagate towards the periphery, in agreement with the behavior illustrated in Fig.~\ref{fig:specialtree}C.
\par
Although our study has been performed for the trees, its results can also be used in the analysis of front propagation in large random Erd\"os-R\'enyi networks. Indeed, it is known \cite{DOR03} that the ER networks are locally approximated by the trees. Hence, if the initial perturbation has been applied to a node and starts to spread over the network, its propagation is effectively taking place on a tree formed by the node neighbors. Previously, we have used this property in the analysis of oscillators entrainment by a pacemaker in large ER networks \cite{KOR04,KOR06}. Only when the activation has already covered a sufficiently high fraction of the network nodes, loops start to  play a role. When this occurs, the activation may arrive at a node along different pathways and the tree approximation ceases to hold. In this opposite situation, a different theory employing the mean-field approximation can however be applied.

\subsection*{Mean-field approximation}

\par
Random ER networks typically have short diameters and diffusive mixing in such networks should be fast. Under the conditions of ideal mixing, the mean-field approximation is applicable; it has previously been used to analyze epidemic spreading \cite{PAS01,col07,COL08}, limit-cycle oscillations and turbulence \cite{nak09}, or Turing patterns \cite{nak10} on large random networks. In this approximation, details of interactions between neighbors are neglected and each individual node is viewed as being coupled to a global mean field which is determined by the entire system. The network nodes contribute to the mean field according to their degrees $k$. The strength of coupling of a node to such global field and also the amount of its contribution to the field are not the same for all nodes and are proportional to their degrees. Thus, a node with a higher number of connections is more strongly affected by the mean field, generated by the rest of the network, and it also contributes stronger to such field. The mean-field approximation is applied below to analyze statistical properties of stationary activation patterns which are well spread over a network and involve a relatively large fraction of nodes.

\par
Similar to publications \cite{nak09,nak10}, we start by introducing the local field 

\begin{equation}
q_{i}=\sum_{j=1}^{N}\!T_{ij}u_{j}
\end{equation}

\noindent determined by the activation of the first neighbors of a network node $i$. Then, the evolution equation \eqref{eq:rdnetT} can be written in the form 

\begin{equation}\label{eq:localfield}
\dot{u}_{i}=f(u_{i}) + D (q_{i} - k_{i} u_{i})\,
\end{equation}

\noindent so that it describes the interaction of the element at node $i$ with the local field $q_{i}$. 
\par
The mean-field approximation consists of the replacement of  the local fields $q_{i}$ by $q_{i} =k_{i}Q$, where the global mean field is defined as

\begin{equation}
Q=\sum_{j=1}^{N}\!w_{j}u_{j}\,.
\end{equation}

\noindent Here, the weights 

\begin{equation}
w_{j}=\frac{k_{j}}{\sum_{n=1}^{N}\!k_{n}}\,
\end{equation} 

\noindent guarantee that the nodes with higher degrees $k_{i}$ contribute stronger to the mean field. After such replacement, Eq.~\eqref{eq:localfield} yields

\begin{equation}\label{eq:mf}
\dot{u}=f(u) + \beta(Q-u)\,
\end{equation}

\noindent where $\beta=Dk$. Note that the index $i$ could be removed because the same equation holds for all network nodes. 
\par
Equation \eqref{eq:mf} describes bistable dynamics of an element coupled to the mean field $Q$. The coupling strength is determined by the parameter $\beta$ which is proportional to the degree $k$ of the considered node. According to Eq.~\eqref{eq:mf}, behavior of the elements located in the nodes with small degrees (and hence small $\beta$) is mostly determined by local bistable dynamics, whereas behavior of the elements located in the nodes with large degrees (and large $\beta$) is dominated by the mean field.

\begin{figure}[t]
\includegraphics{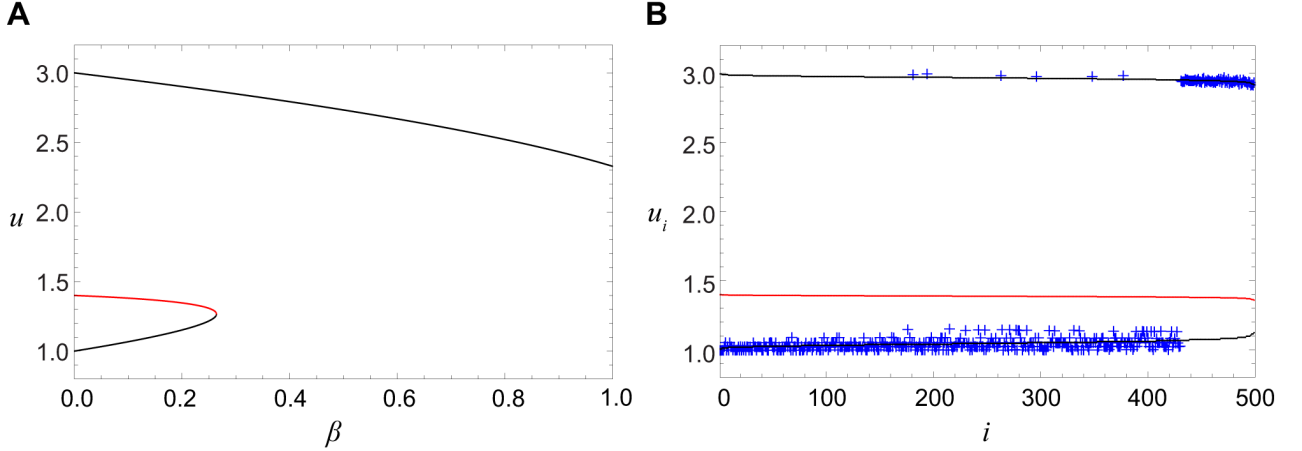}
\caption{{\bf The stationary activity pattern and the mean-field bifurcation diagram}. (A) The bifurcation diagram of Eq.~\eqref{eq:mf} for the mean field $Q=1.5$. (B) Activity distribution in the stationary pattern in the ER network of size $N=500$ and mean degree $\langle k\rangle=7$ at $D=0.01$ is compared with the activator levels $u$ predicted by the mean-field theory for $Q=1.5$. Blue crosses show the simulation data. Black  and red curves indicate stable  and unstable fixed points of the mean-field equation \eqref{eq:mf}. The other parameters are $r_{1}=1,r_{2}=1.4,r_{3}=3$.}  
\label{fig:mf}
\end{figure}

\par
The fixed points of  Eq.~\eqref{eq:mf} yield activator levels $u$ in single nodes coupled with strength $\beta$ to the mean field $Q$. Self-organized stationary patterns on a random ER network can be analyzed in terms of this mean-field equation. Indeed, the activator level in each node $i$ of a pattern can be calculated from Eq.~\eqref{eq:mf}, assuming that the node is coupled to the mean field determined by the entire network. In Fig.~\ref{fig:mf}, the mean-field approximation is applied to analyze the stationary pattern shown in Fig.~\ref{fig:ERstationary}. This pattern has developed in the ER network of size $N=500$ and mean degree $\langle k\rangle=7$ when the diffusion constant was fixed at $D=0.01$. The mean field corresponding to such pattern was computed in direct numerical simulations and is equal to $Q=1.5$. Substituting this value of $Q$ into Eq.~\eqref{eq:mf}, activator levels $u$ in single node, coupled to this mean field can be obtained. In Fig.~\ref{fig:mf}A, the activator level $u$ is plotted as a function of the parameter $\beta$. When a node is decoupled ($\beta = 0$), Eq.~\eqref{eq:mf} has three fixed points $r_{1},r_{2},r_{3}$. As $\beta$ is increased, the system undergoes a saddle-node bifurcation beyond which only one stable fixed point remains. 
\par
According to the definition of the parameter $\beta$, each node $i$ with degree $k_i$ is characterized by its own value $\beta_i=Dk_i$ of this bifurcation parameter. Therefore, the fixed points of Eq.~\eqref{eq:mf} can be used to determine the activation levels for each node $i$, if its degree $k_{i}$ is known. The stationary activity distributions, predicted by the mean-field theory and found in direct numerical simulations, are displayed in Fig.~\ref{fig:mf}B, where the nodes are ordered according to their increasing degrees. Note that the value $Q=1.5$ of the mean field, used to determine the activity levels, has been taken here from the numerical simulation. As we see, data points indeed lie on the two stable branches of the bifurcation diagram, indicating a good agreement with the mean-field approximation. In the Supporting Information S1, a similar mean-field analysis is performed for self-organized stationary activity patterns on scale-free networks.

\section*{Discussion}

\par
Traveling fronts represent classical examples of non-equilibrium patterns in bistable reaction-diffusion media. As shown in our study, such patterns are also possible in networks of diffusively coupled bistable elements, but their properties are significantly different. In addition to spreading or retreating activation fronts, stationary fronts are found within large parameter regions. The behavior of the fronts is highly sensitive to network architecture and degrees of network nodes play an important role here.
\par
In the special case of regular trees, an approximate analytical theory could be constructed. The theory reveals that branching factors of the trees and, thus, the degrees of their nodes, are essential for front propagation phenomena. By using this approach, front pinning conditions could be derived and parameter boundaries, which separate pinned and traveling fronts, could be determined. As we have found, propagation conditions are different for the fronts traveling from the tree root to the periphery or in the opposite direction. Generally, all fronts become pinned as the diffusion constant is gradually reduced. While the theory has been developed for regular trees, where the branching factor is fixed, it is also applicable to irregular trees where node degrees are variable. Indeed, at sufficiently weak diffusion the front pinning occurs locally and its conditions are effectively determined only by the degrees of the nodes at which a front becomes pinned. 
\par
The results of such analysis are relevant for understanding the phenomena of activation spreading and pinning in large random networks. It is well known (see, e.g., \cite{DOR03}) that, in the large size limit, random networks are locally approximated by the trees. If the number of connections (the degree) of a node is much smaller than the total number of nodes in a network, the probability that a neighbor of a given node is also connected to another neighbor of the same node is small, implying that the local pattern of connections in the vicinity of a node has a tree structure. This property holds as long as the number of nodes in the considered neighborhood is still much smaller that the total number of nodes in the network. Previously, the local tree approximation has been successfully used in the analysis of pacemakers in large random oscillatory networks \cite{KOR04,KOR06}.
\par
When activation is applied to a node in a large random network, it spreads through a subnetwork of its neighbors and, at sufficiently short distances from the original node, such subnetwork should be a tree. Hence, our study of front propagation on the trees is also providing a theory for the initial stage of front spreading from a single activated node in large random networks. Depending on the diffusion constant and other parameters, the fronts may become pinned while the activation has not yet spread far away from the original node. Whenever this takes place, the approximate pinning theory, constructed for the trees, is applicable. 
\par
On the other hand, if the activation spreads far from the origin and a large fraction of network nodes become thus affected, the patterns can be well understood with the mean-field approximation. This approximation, proposed in the analysis of infection spreading on networks \cite{PAS01}, has also been applied to analyze Turing patterns in network-organized activator-inhibitor systems \cite{nak10} and effects of turbulence in oscillator networks \cite{nak09}. In this paper, we have applied this approximation to the analysis of stationary activity distributions in random Erd\"os-R\'enyi and scale-free networks of diffusively coupled bistable elements. We could observe that, within the mean-field approximation, statistical properties of network activity distributions are well reproduced. It should be noted that, similar to previous studies \cite{nak09,nak10}, the mean-field values used in the theory were taken from direct numerical simulations and were not obtained through the solution of a consistency equation. Hence, we could only demonstrate that such an approximation is applicable for the statistical description of the emerging stationary patterns, but did not use it here for the prediction of such patterns. 
\par
Thus, our investigations have shown that a rich behavior involving traveling and pinned fronts is characteristic for networks of diffusively coupled bistable elements. In the past, pinned fronts were observed in the experiments using weakly coupled bistable chemical reactors on a ring \cite{LAP92,BOO94}. It will be interesting to perform similar experiments for the networks of coupled chemical reactors. Recent developments in nanotechnology allow to design chemical reactors at the nanoscale and couple them by diffusive connections to build networks \cite{kar01}. It should be also noted that, while the chemical Schl\"ogl model has been used in our numerical simulations, the results are general and applicable to any networks formed by diffusively coupled bistable elements. The phenomena of front spreading and pinning should be possible for diffusively coupled ecological populations and similar effects may be involved when epidemics spreading under bistability conditions is considered.

\section*{Methods}

{\bf Bistable dynamics}. 
The Schl\"ogl model \cite{SCHLOEGL72} corresponds to a hypothetical reaction scheme

\begin{eqnarray}
A+2X & \overset{c_1}{\underset{c_2}{\rightleftarrows}} & 3X\nonumber\\
X & \overset{c_3}{\underset{c_4}{\rightleftarrows}} & B\,.
\end{eqnarray}

\noindent If concentrations of reagents $A$ and $B$ are kept fixed, the rate equation for the concentration $u$ of the activator species $X$ reads

\begin{equation}\label{eq:rate1}
\dot{u}(t) = - c_{2} u^{3}(t) + c_{1} a u^{2}(t) -c_{3} u(t) + c_{4} b
\end{equation}

\noindent where the coefficients $c_1,c_2,c_3,c_4$ are rate constants of the reactions; $a=[A]$, $b=[B]$ and $u=[X]$ are concentrations of chemical species. By choosing appropriate time units, we can set $c_{2}=1$. Then, the right side of Eq.~\eqref{eq:rate1}, can be written as

\begin{equation}\label{eq:localdynamics}
f(u)=-(u-r_{1})(u-r_{2})(u-r_{3}) 
\end{equation}

\noindent where the parameters $r_{1},r_{2},r_{3}$ satisfy the conditions 

\begin{eqnarray}
c_{1} &=& \frac{r_{1}+r_{2}+r_{3}}{a}\nonumber\\ 
c_{3} &=& r_{1}r_{2} + r_{1}r_{3} + r_{2}r_{3}\nonumber\\ 
c_{4} &=& \frac{r_{1}r_{2}r_{3}}{b}\,.
\end{eqnarray}

\par 
The cubic polynomial $f(u)$ has three real roots which correspond to the steady states (fixed points)  of the dynamical system~\eqref{eq:rate1}. 

{\bf Networks}.
Erd{\"o}s-R{\'e}nyi networks were constructed by taking a large number $N$ of nodes and randomly connecting any two nodes with some probability $p$. This construction algorithm yields a Poisson degree distribution with the mean degree $\langle k \rangle=pN$ \cite{alb02}. In our study we have considered the largest connected component network, namely, we have removed the nodes with the degree $k=0$. 
\par
Tree networks with branching factor $k-1$ were constructed by a simple iterative method. We start with a single root node and at each step add $k-1$ nodes to each existing node with the degree $k=1$. After $L$ steps this algorithm leads to a tree network with the size $N=\sum_{l=1}^{L}(k-1)^{l-1}$, where the root node has degree $k-1$, the last added nodes have degree $1$ and all other nodes have degree $k$. In our numerical simulations we have also used complex trees consisting of component trees with different fixed branching factors which are connected at their origins. 
\par
Scale-free networks, considered in the Supporting Information S1, were constructed by the preferential attachment algorithm of Bar{\'a}basi and Albert \cite{alb02}. Starting with a small number of $m$ nodes with $m$ connections, at each next time step a new node is added, with $m$ links to $m$ different previous nodes. The new node will be connected to a previous node $i$, which has $k_{i}$ connections, with the probability $k_{i}/\sum_{j}\!k_{j}\,$. After many time steps, this algorithm leads to a network composed by $N$ nodes with the power-law degree distribution $P(k) \sim k^{-3} $ and the mean degree $\left<k\right>=2m$.
\par
To display the networks in Figs.~\ref{fig:ERstationaryNeato}, \ref{fig:specialtree} and S2B we have used the Fruchterman-Reingold force-directed algorithm which is available in the open-source Python package NetworkX \cite{networkx}. This network visualization algorithm places the nodes with close degrees $k$ near one to another in the network projection onto a plane. 

{\bf Numerical methods}.
For networks of coupled bistable elements, simulations were carried out by numerical integration of Eq. \eqref{eq:rdnetL} using the explicit Euler scheme 

\begin{equation}
u_{i}^{(t+1)}=u_{i}^{(t)}+ dt\left[f\left(u_{i}^{(t)}\right)+D\sum_{j=1}^{N}\!L_{ij}u_{j}^{(t)}\right]
\end{equation}

\noindent with the time step $dt=10^{-3}$. The integration was performed for $5\times10^{5}$ steps. The initial conditions were $u_{i}=1$ for all network nodes $i$, except a subset of nodes to which initial activation was applied and where we had $u_{i}=3$. The explicit Euler scheme with the same time step $dt$ was also used to integrate Eq.~\eqref{eq:kchain} which describes patterns on regular trees. 

%=======================================================================================================
%
%
\section*{Supporting Information}

{\bf Supporting Information S1}. 
The results of the numerical simulations of the bistable Schl\"ogl model~\eqref{eq:rdnetL} for scale-free networks are provided. Traveling fronts and stationary localized patterns are reported for networks with mean degree $\langle k \rangle=6$ and sizes $N=150$ or $N=500$ nodes. The observed stationary pattern is compared with the mean-field bifurcation diagram.

\section*{Acknowledgments}
Financial support from the DFG Collaborative Research Center SFB910 ``Control of Self-Organizing Nonlinear
Systems'' and from the Volkswagen Foundation in Germany is gratefully acknowledged.

\section*{Author Contribution}
Designed the study: NK ASM. Performed the simulations: NK. Conceived the analytical approximations: NK HK ASM. Wrote the article: NK HK ASM.

\newpage

\section*{Supporting Information S1}

In this supporting information, results of numerical simulations of the bistable Schl\"ogl model~(3) for scale-free networks with mean degree $\langle k \rangle=6$ and sizes $N=150$ or $N=500$ are reported. The model~(3) with the parameters $r_{1}=1$ and $r_{3}=3$ is chosen; the parameter $r_{2}$ and the diffusive constant $D$ were varied in the simulations.
\par
Both traveling and pinned fronts were observed. To initiate a traveling front, a node at the periphery with the degree $k=3$ was set into the active state $r_3$, whereas the rest of the nodes were in the passive state $r_1$. This initial configuration generated a front which spread over the entire network. 
\par
Front propagation is seen in Fig.~S1, where the nodes are grouped according to their distance from the first activated node and the average value $\rho_h$ of the activator density $u$ in each group is plotted as a function of the distance $h$. Three snapshots of the traveling front at different times are displayed. At $t=0$, the activation is localized on one node. By time $t=10$, it spreads to the second neighbors of the original node. At $t=21$, the activation extends to the fifth neighbors of the original node, covering almost the entire network. Note that a definite traveling front is observed only while the activation is still close to the origin. At the final stage, the front rapidly broadens and the transition to the final uniform active state is quickly taking place.

\begin{figure}[h!]
\includegraphics[width=\textwidth]{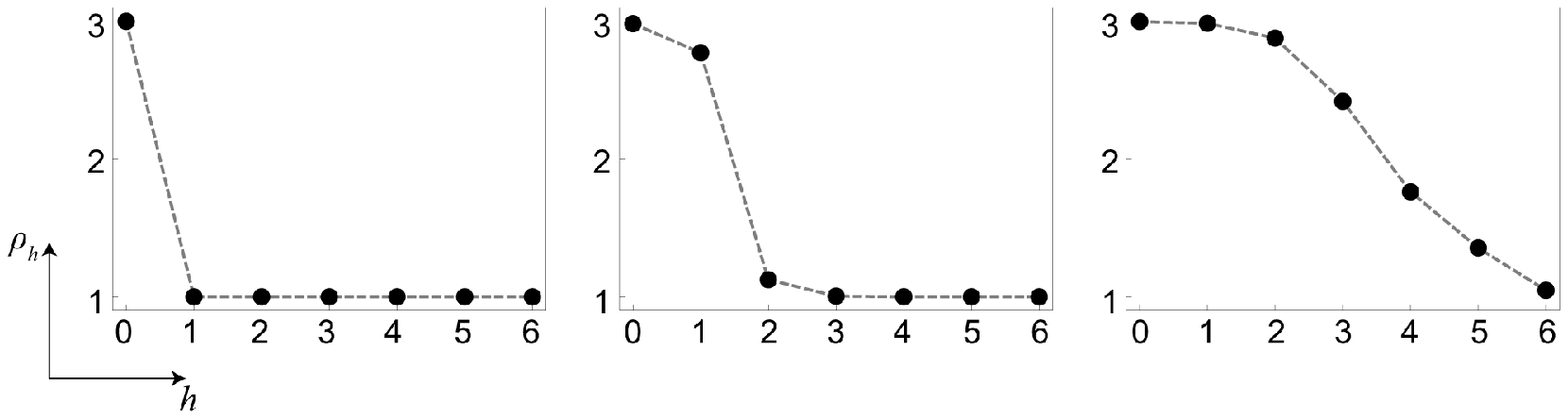}
\caption*{{\bf Figure S1.} Activation front in a scale-free network with mean degree $\langle k \rangle=6$ and size $N=500$. Three consequent snapshots of activity patterns at times $t=0,10,21$ are displayed. Quantity $\rho_{h}$ is the average value of the activator density $u$ in the subset of network nodes located at distance $h$ from the node which was initially activated. The model parameters are $r_1=1,r_2=1.2,r_2=3$ and the diffusion constant is $D=0.1$.}  
\end{figure}

\par
When one of the hub nodes was initially activated, a spreading activation front could not be produced. In this case, retracting fronts were observed, if a compact group of nodes with large degrees was initially activated. For weak diffusive coupling, stationary localized patterns were also found, either by appropriate choosing initial conditions, or when traveling fronts was getting pinned at some nodes so that the spreading activation could not reach all network nodes. Two examples of such stationary patterns are shown in Fig.~S2. 

\begin{figure}[t!]
\includegraphics[width=\textwidth]{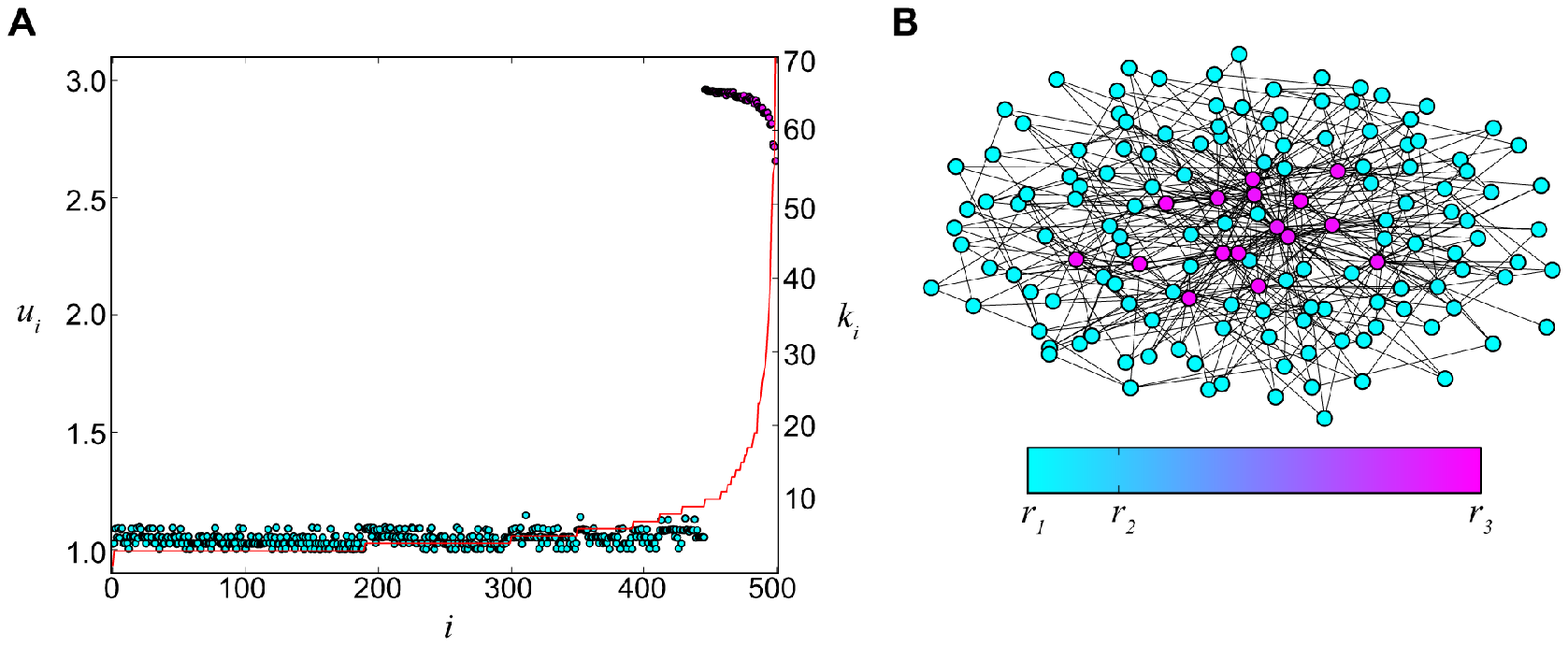}
\caption*{{\bf Figure S2.} (A) Dependence of the activation level $u_{i}$ on the degrees $k_{i}$ of the nodes $i$ for a stationary pattern in the scale-free network of size $N=500$ and mean degree $\langle k \rangle=6$. The red curve shows the degrees of the nodes. (B) Stationary pattern in the scale-free network of size $N=150$ and mean degree $\langle k \rangle=6$. The nodes with higher degrees are located closer to the center. The nodes are colored according to their activation level, as indicated in the bar. The parameters are $r_{1} = 1, r_{2} = 1.4$ and $r_{3} = 3$; the diffusion constant is $D = 0.01$.}  
\end{figure}

\begin{figure}[h!]
\includegraphics{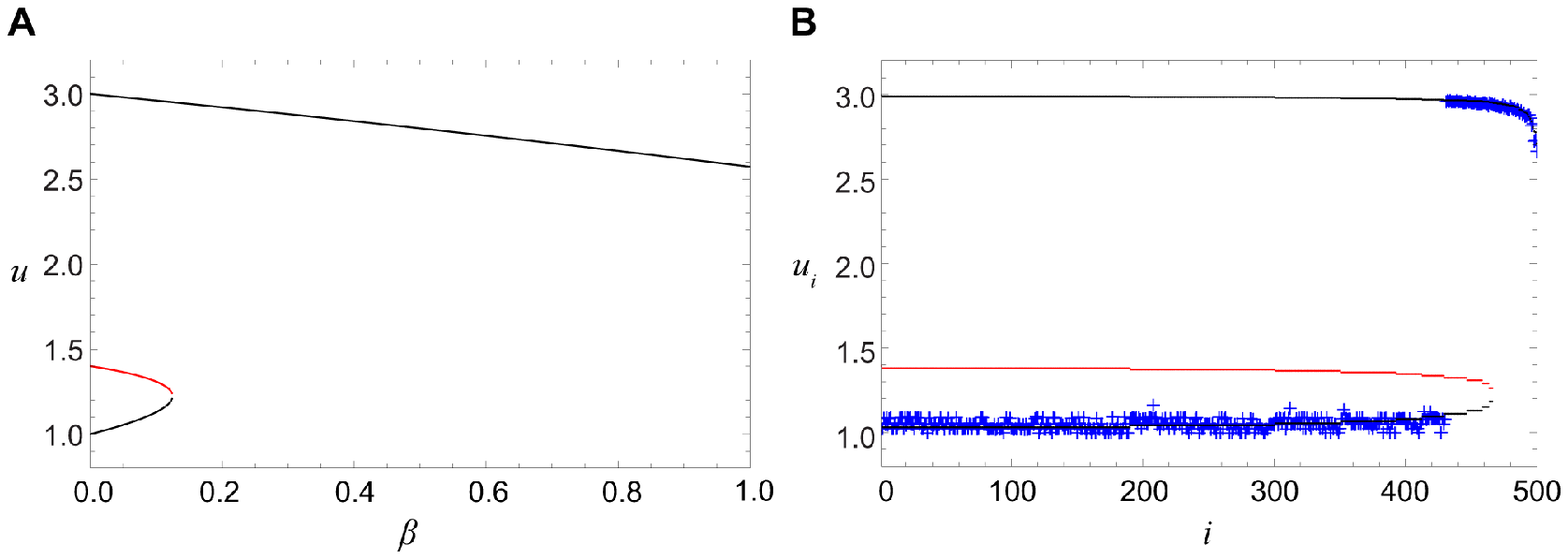}
\caption*{{\bf Figure S3.} (A) The bifurcation diagram of Eq. (16) for the mean field $Q=1.68$. (B) Activity distribution in the stationary pattern in the scale-free network of size $N=500$ and mean degree $\langle k\rangle=6$ at $D=0.01$ is compared with the activator levels $u$ predicted by the mean-field theory for the mean field $Q=1.68$ of the numerically computed pattern. Blue crosses show the simulation data. Black  and red curves indicate stable  and unstable fixed points of the mean-field equation (16). The other parameters are $r_{1}=1,r_{2}=1.4,r_{3}=3$.}  
\end{figure}

\par
The mean-field approximation could be used to describe self-organized stationary patterns on the scale-free networks. The mean-field computed in the numerical simulations for the stationary pattern shown in Fig.~S2A was equal to $Q=1.68$. Substituting this value into the Eq.~(16) we obtain the bifurcation diagram of a single node coupled to this mean-field. The stationary pattern is compared with the mean-field bifurcation diagram in Fig.~S3B. The curves are predictions of the mean-field approximation and the crosses show the simulation data. We see that the data points are distributed along the two stable branches of the bifurcation diagram, indicating good agreement with the mean-field approximation.

%=======================================================================================================
\end{document}